\documentclass[conference]{ieeeconf}
\IEEEoverridecommandlockouts

\usepackage{amsmath,amssymb,amsfonts,amsthm,bbm}
\usepackage{graphicx}
\usepackage{comment}
\usepackage{textcomp}
\usepackage[makeroom]{cancel}
\usepackage{indentfirst}
\usepackage{ulem}
\usepackage{tikz}
\usepackage{epstopdf} 
\usepackage{epsfig}
\usepackage[dvipsnames]{xcolor}
\usepackage{algorithm}
\usepackage{algorithmic}
\usepackage{hyperref}
\usepackage{etoolbox}

\usepackage{optidef}

\DeclareMathOperator*{\argmin}{arg\,min}

\usepackage{mathtools}
\mathtoolsset{showonlyrefs}

\usepackage[utf8]{inputenc}
\usepackage[english]{babel}
\usepackage[noadjust]{cite}

\renewcommand{\emph}{\textit}

\newbool{short}

\boolfalse{short} 

\newtheorem{theorem}{Theorem}
\newtheorem{corollary}{Corollary}
\newtheorem{assumption}{Assumption}
\newtheorem{lemma}{Lemma}
\newtheorem{definition}{Definition}
\newtheorem{problem}{Problem}
\newtheorem{proposition}{Proposition}

\usepackage{csquotes}
\usepackage{float}
\usepackage{accents}

\def\BibTeX{{\rm B\kern-.05em{\sc i\kern-.025em b}\kern-.08em
    T\kern-.1667em\lower.7ex\hbox{E}\kern-.125emX}}

\newcommand{\red}[1]{\textcolor{red}{#1}}

\newcommand{\R}{\mathbb{R}}

\newcommand{\Z}{\mathbb{Z}}

\newcommand{\n}[1]{\text{#1}}

\title{\LARGE \bf
Computationally Constrained
First-Order Prediction-Correction Approach for Time-Varying Convex Optimization
}

\author{Negin Musavi$^{1}$, Oscar Jed R. Chuy$^{2}$, Melkior Ornik$^{1}$, Matthew T. Hale$^{2}$
\thanks{
    $^{1}$Department of Aerospace Engineering, University of Illinois Urbana-Champaign, Champaign, IL, USA. 
    Emails: \texttt{\{nmusavi2,mornik\}@illinois.edu}.
    }%
\thanks{
    $^{2}$School of Electrical and Computer Engineering, Georgia Institute of Technology, Atlanta, GA, USA. 
    Emails: \texttt{\{ochuy3,matthale\}@gatech.edu}.
    }%
\thanks{Chuy and Hale were supported by the Office of Naval Research under grant N00014-26-1-2068, and 
Musavi and Ornik were supported by the Office of Naval Research under grant N00014-25-1-2519 and N00014-25-1-2369.}%
}

\begin{document}

\maketitle
\thispagestyle{empty}
\pagestyle{empty}

\begin{abstract}
    Time-varying optimization methods often predict the locations of time-varying optima, then sample the objective function being minimized, and then compute corrections to those predictions.     
    However, existing methods typically do not account for settings in which the number of computations is limited. 
    Therefore, in this work we formulate a prediction-correction algorithm that is designed to obey a budget constraint on the total number of iterations it can execute over a given time horizon.
    We first identify a threshold on the average number of computations between samples of the objective, above which a bound from prior work certifies contraction of iterates toward
    a minimizer. 
    Below this threshold, we introduce a finite-horizon contraction certificate, which we use to give budget conditions under which the tracking error admits an explicit bound that decays at a prescribed rate.
    We also present an algorithm to determine how many prediction and correction steps should be executed for each sample of the objective, while obeying a given budget on the allowable number of computations.
    Empirically, we show in a simulation that 
    our approach reduces cumulative tracking error by~{$45.5\%$} relative to a comparable baseline approach. 
\end{abstract}
\section{Introduction}\label{sec:introduction}
Time-varying optimization problems 
arise in robotics~\cite{SimonettoP-Cmethods_TVConvexOP_2016, SimonettoPCaTVcop2017}, power systems~\cite{LinOPCausLEtvo2026}, transportation networks~\cite{Simonettotvcopt-saa2020}, communication systems~\cite{BianchinTVopLTIPDgf2022}, and distributed systems~\cite{SimonettoDPCmLCtvcp2019,SimonettoDPCmntvcop}. 
A common way to solve such problems is with a prediction-correction method~\cite{SimonettoPCmethodsfortvcop2015}, which uses 
the objective sampled at a prior time 
to predict how the solution will evolve in the future.
Then, after resampling the objective, corrections are made 
to this prediction 
to account for differences 
between the predicted and actual movements of the solution. 
Such methods often use gradient or Newton steps to correct for those differences~\cite{SimonettoPCaTVcop2017,SimonettoP-Cmethods_TVConvexOP_2016, SimonettoPCmethodsfortvcop2015}.  
After completing some number of correcting steps,
the prediction-correction cycle is repeated by predicting the subsequent evolution of the solution, then sampling the objective, and then correcting those predictions. 
This class of method has been shown to track minimizers with error that is acceptable in a wide range of applications~\cite{PaternainPCmMPC2018,SimonettoDPCmntvcop, SimonettoDPCmLCtvcp2019}.

Prior works such as~\cite{SimonettoP-Cmethods_TVConvexOP_2016, SimonettoPCmethodsfortvcop2015, PaternainPCmMPC2018, SimonettoDPCmntvcop, SimonettoPCaTVcop2017, SimonettoDPCmLCtvcp2019, LinOPCausLEtvo2026, BASTIANELLOEvPCmTVcop2023} 
have used prediction-correction methods in applications including target tracking, model-predictive control, and various decentralized formulations. 
All of these works (and  many others) use fixed numbers of prediction steps and correction steps between samples of the objective. 
However, in practice it might not be possible to execute the desired number of computations, e.g., due to constraints on 
resources such as 
battery power, which can limit the total number of computations that can be executed over a given time horizon. 
Such constraints have driven interest in various forms of resource awareness across different settings. 
For example, existing works in~\cite{ScokaertSOMPC1999,LiEAMAKhpPPOMODN2025,EqtamiETcDTs2010, seewaldEAesceMAsBC2024, ZhangEAMUAVctsRLBEC2026,ZhangEAOpCAEVcVTn2024, KimEALPCAEVUTDVitlv2025,ZhangETNMPCEAUAVoa2026} 
have investigated resource awareness in multi-agent control, connected and autonomous electric vehicles, and event-triggered control, 
but to the best of our knowledge there has not been a systematic study of time-varying optimization under computational constraints. 

Therefore, in this paper, we develop a computationally constrained prediction-correction algorithm for time-varying optimization.
Computation constraints are encoded as an upper bound on the total number of prediction
and correction steps that can be executed over a given time horizon, which we
call the ``budget'' for a problem. 
We consider a first-order prediction-correction algorithm that uses gradient
steps for both predictions and corrections. 

We first identify a critical threshold on the number of computations 
that can be executed for each sample of the objective function.
Then we show that if the budget permits at least that many computations to be executed
for each sample of the objective, 
then we can certify that the iterates of the first-order algorithm
contract toward a minimizer.

In contrast, if the threshold we derive cannot be met for some samples
of the objective, then 
the aforementioned contraction guarantee no longer applies. 
However, we show that one can still allocate prediction and correction steps 
so that the distance to the minimizer admits a
different form of 
explicit bound that decreases at a prescribed rate. 
Then, we develop an algorithm for determining how many prediction and correction steps should be completed for each sample of the objective function, and this algorithm is designed to attain the error bounds we derive on the distance to a time-varying optimum.

\subsection{Summary of Contributions}
The contributions of this paper are:
\begin{itemize}
    \item We identify a critical threshold on the average number of computations per sample of the objective.
    Above this threshold, we 
    guarantee that a bound on the distance to the minimizer converges geometrically to a neighborhood whose size is proportional 
    to the time that elapses between taking samples of the objective (Corollary~\ref{thm:regime1}).
    \item Below this threshold, we introduce a finite-horizon contraction certificate that 
    encodes
    a desired rate of decay in the distance from iterates to a minimum. Then we derive budget conditions under which this certificate can be satisfied and thus ensure that
    the distance to a minimizer decreases at a user-specified rate 
    (Theorems~\ref{thm:exact-feasibility} and~\ref{thm:exact-tracking}).
    \item We develop an algorithm that determines how many prediction and correction steps to execute for each sample 
    of the objective while obeying
    a total computation budget constraint
    (Algorithm~\ref{alg:regime2}, Theorems~\ref{thm:exact-tracking} and~\ref{thm:uniform-budget}).
    \item We show in simulation that our method tracks time-varying optima
    with {$45.5\%$} less cumulative error than
    a comparable baseline approach~(Section~\ref{sec:simulations}). 
\end{itemize}

The rest of this paper is organized as follows. 
Section~\ref{sec:problem} provides background and a problem formulation. 
Section~\ref{sec:pcmethod} develops the computationally constrained algorithm, and Section~\ref{sec:pctracking} computes error bounds.
Then, Section~\ref{sec:simulations} presents simulations, and~Section~\ref{sec:conclusion} concludes the paper.

\subsection{Notation}
We use~$\R$ to denote the reals and~$\R_{\geq0}$ to denote the non-negative reals.
For an integer~$k$, 
we use~$\Z_{\geq k}$ to denote the set of integers greater than
or equal to~$k$. 
We denote the~$2$-norm of a vector~$x \in \R^n$ by~$\|x\|$. 
We use~$I$ to denote the identity matrix. 
For a function~$f:\R^{n}\times \R_{\geq 0} \rightarrow \R$
that takes arguments~$(x, t)$,
we denote its gradient with respect to~$x$ by~$\nabla_{x}f$, and we denote 
its derivative with respect to~$t$ by~$\nabla_{t}f$.
We use~$\nabla_{xx}f$ to denote the Hessian of~$f$ with respect to~$x$, and~$\nabla_{tx}f$ is the 
time derivative of~$\nabla_{x}f$. 

\section{Preliminaries and Problem Formulation} \label{sec:problem}
In this section, we first define the time-varying optimization problem in Section~\ref{subsec:tvop}. Then in Section~\ref{subsec:pcmethod}, we describe the prediction--correction method. Finally, in Section~\ref{ss:prob}, we formulate the problem that we address in this paper. 

\subsection{Time-Varying Optimization} \label{subsec:tvop}

Let $f : \mathbb{R}^n \times \mathbb{R}_{\geq 0} \to \mathbb{R}$ be a time-varying function, and let $T_c > 0$ be a finite time horizon.
We make the following assumption on $f$.

\begin{assumption}\label{ass:regularity}
    The function~$f$
    is twice continuously differentiable in~$(x,t)$ and there exist constants~$0 < m \leq L$ such that
    \begin{equation} \label{eq:hessian_bounds}
        mI
        \preceq
        \nabla_{xx}f(x,t)
        \preceq
        LI 
        \quad
        \n{for all } x\in \R^{n},~t\in [0,T_{c}], 
    \end{equation}
    as well as a constant~$\Gamma>0$ such that
    \begin{equation} \label{eq:global_temporal_bound}
        \|\nabla_{tx}f(x,t)\|
        \leq
        \Gamma
        \quad
        \n{for all } x\in \R^{n},~t\in [0,T_{c}].
    \end{equation}
\end{assumption}
Assumption~\ref{ass:regularity} ensures that~$f(\cdot,t)$ is uniformly strongly convex and smooth, which guarantees that it has a unique minimizer for each
$t \in [0, T_c]$.
The inequalities in~\eqref{eq:hessian_bounds} provide uniform curvature bounds over time, 
and the constant~$\Gamma$ bounds the temporal variation of the gradient. 
The bound in~\eqref{eq:global_temporal_bound} means that the time derivative of~$\nabla_x f(x, t)$ is bounded.

Under Assumption~\ref{ass:regularity}, consider time-varying convex optimization
problems of the form
\begin{equation} \label{eq:tv_problem}
  x^\star(t) := \argmin_{x \in \mathbb{R}^n} f(x, t),
  \qquad t \in [0, T_c],
\end{equation}
where $x \in \mathbb{R}^n$ is a decision variable. The objective~$f(\cdot,t)$ is only known at past and current times~$t$, 
and future objectives are not known \emph{a priori}. 

Suppose that the objective can only be accessed at discrete time instances.
Then, the problem~\eqref{eq:tv_problem} can be considered in discrete time.
Following~\cite{nesterov2018lectures, boyd2004convex}, the objective
is sampled at the time instances~$t_{k} := kh$, where~$h>0$ is a constant
sampling interval and~$k\in\{0,1,\ldots,T\}$ for a finite time
horizon~$T\in\Z_{\geq 1}$ such that~$Th\leq T_{c}$. This yields the sequence
of problems
\begin{equation}\label{eq:Dtv_optimization_problem}
    x^\star(t_{k}) := \argmin_{x\in \R^{n}} f(x,t_{k}),
    \quad k = 0,1,\ldots,T,
\end{equation}
which is the problem considered in the remainder of this paper. At
each~$t_{k}$, the current objective~$f(\cdot,t_{k})$ and its derivatives
at~$t_k$, including~$\nabla_{xx}f(\cdot,t_k)$
and~$\nabla_{tx}f(\cdot,t_k)$, are known.


\subsection{Prediction-Correction Method} \label{subsec:pcmethod}

In this section we describe the prediction--correction method
of~\cite{SimonettoPCaTVcop2017}, which proceeds as follows. Starting with an initial estimate~$x_{0}\in\R^n$, the method  generates estimates~$\{x_{k}\}_{k=1}^{T}$ of the 
optimal trajectory~$\{x^\star(t_k)\}_{k=1}^{T}$ in~\eqref{eq:Dtv_optimization_problem}. For
$k \in \{0, 1, \ldots, T-1\}$, the method uses the estimate $x_k$ of
$x^\star(t_k)$ to predict $x^\star(t_{k+1})$ by executing $P_k$ prediction
steps. Then, the objective $f(\cdot, t_{k+1})$ is sampled, and $C_k$
correction steps are executed to correct the prediction. The final
correction step produces $x_{k+1}$, from which this process repeats. The
prediction and correction steps are described in
Sections~\ref{subsec:predict_theory} and~\ref{subsec:correct_theory}.
In~\cite{SimonettoPCaTVcop2017}, the numbers of prediction and correction
steps are the same for every $k$; here, they may vary with $k$.

\subsubsection{Prediction} \label{subsec:predict_theory}
Between two sampling times $t_k$ and $t_{k+1}$, $P_k$ prediction steps are
executed to predict $x^\star(t_{k+1})$ from $x_k$. A first-order expansion
in space and time gives
\begin{equation*}
  \nabla_x f(z, t_{k+1}) \approx g_k + H_k (z - x_k) + h v_k,
\end{equation*}
where $g_k := \nabla_x f(x_k, t_k)$, $H_k := \nabla_{xx} f(x_k, t_k)$,
and $v_k := \nabla_{tx} f(x_k, t_k)$. Since $f(\cdot, t_{k+1})$ is not
known before it is sampled, solving $\nabla_x f(z, t_{k+1}) = 0$ exactly
may not be practical. Instead, as in~\cite{SimonettoPCaTVcop2017}, a
relaxed optimality condition is imposed by seeking a point $z$ that
satisfies $\nabla_x f(z, t_{k+1}) = (1 - \gamma) g_k$ with
$\gamma \in [0, 1]$. This relaxation leads to the prediction step
\begin{equation} \label{eq:prediction_iteration}
  x_k^{p, \ell+1} = x_k^{p, \ell}
  - \alpha \left[ H_k \left( x_k^{p, \ell} - x_k \right)
  + h v_k + \gamma g_k \right]
\end{equation}
for $\ell \in \{0, \ldots, P_k - 1\}$, where $x_k^{p, \ell}$ is the
$\ell$th prediction iterate computed for the $k$th sample of the objective,
$x_k^{p, 0} = x_k$, and $\alpha \in (0, \frac{2}{L})$ is the prediction step size.

\subsubsection{Correction} \label{subsec:correct_theory}
After all~$P_k$ prediction steps are completed, the objective
function~$f(\cdot,t_{k+1})$ is sampled, and~$C_k$ correction steps are
executed. Initialized at~$x_{k+1}^{c,0}=x_{k}^{p,P_{k}}$, the correction step is
\begin{equation}\label{eq:correction_iteration}
    x_{k+1}^{c,\ell+1}
    =
    x_{k+1}^{c,\ell}
    -
    \beta\nabla_x f\big(x_{k+1}^{c,\ell},t_{k+1}\big)
\end{equation}
for~$\ell\in\{0,\dots,C_k-1\}$, where~$x_{k+1}^{c, \ell}$ is
the~$\ell^{\n{th}}$ correction iterate computed for
the~$(k+1)^{\n{th}}$ sample of the objective, and~$\beta \in (0, \frac{2}{L})$
is the correction step size. After~$C_{k}$ correction steps,
$x_{k+1}:= x_{k+1}^{c,C_{k}}$, and then this process of prediction and
correction repeats.


\subsection{Problem Formulation} \label{ss:prob}

\begin{definition} \label{def:allocation}
    An allocation is a collection~$\{(P_k, C_k)\}_{k=0}^{T-1}$, with
    $P_k, C_k \in \Z_{\geq 0}$, of the numbers of prediction and correction
    steps~\eqref{eq:prediction_iteration}--\eqref{eq:correction_iteration}
    executed between consecutive sampling times $t_k$ and $t_{k+1}$.
\end{definition}

The sum~$K_k := P_k + C_k$ is the number of iterations used to
compute~$x_{k+1}$ from~$x_k$. For a given allocation $\{(P_k, C_k)\}_{k=0}^{T-1}$, let
$\{x_k\}_{k=0}^{T}$ denote the iterates generated
by~\eqref{eq:prediction_iteration}--\eqref{eq:correction_iteration} from an initial estimate
$x_0$. We define the tracking error of this allocation as
\begin{equation} \label{eq:tracking_error}
  e_k := \| x_k - x^\star(t_k) \|, \qquad k = 0, \ldots, T.
\end{equation}

We consider a total iteration budget $B \in \mathbb{Z}_{\geq 1}$ over the
$T$ sampling intervals. Each prediction iteration~\eqref{eq:prediction_iteration} and
each correction iteration~\eqref{eq:correction_iteration} incurs one unit of cost,
while sampling the objective and evaluating $g_k$, $H_k$, and $v_k$ do not
count toward the budget. Our goal is to solve the following problem.

\begin{problem} \label{prob:main}
Given an iteration budget $B$, determine an allocation
$\{(P_k, C_k)\}_{k=0}^{T-1}$ with bounded tracking errors $e_k$ for all
$k \in \{1, \ldots, T\}$ such that
\begin{equation} \label{eq:total_budget_constraint}
  \sum_{k=0}^{T-1} \left( P_k + C_k \right) \leq B.
\end{equation}
\end{problem}

The budget constraint permits the per-sample totals $K_k := P_k + C_k$ to
vary across sampling intervals. To address Problem~\ref{prob:main}, we first
characterize how the allocation affects the tracking error. We then use this
characterization to develop allocation rules that meets the total iteration
budget and admit a finite-horizon tracking bounds.

\section{Computationally Constrained Prediction-Correction Method} \label{sec:pcmethod}

In this section, we develop a computationally constrained allocation strategy for
prediction and correction iterations 
that 1)~provides a finite-horizon tracking guarantee and 2)~satisfies the budget constraint in~\eqref{eq:total_budget_constraint}.
We first present tracking error
bounds for a general allocation~$\{(P_{k},C_{k})\}_{k=0}^{T-1}$. Then,
given a budget~$B$, we derive allocations under two different regimes.

\subsection{Tracking-Error Bound and Budget Regimes}
Let
\begin{align}\label{eq:rho_pc}
    \rho_p
    &:=
    \max\left\{|1-\alpha m|,|1-\alpha L|\right\},\\
    \rho_c
    &:=
    \max\left\{|1-\beta m|,|1-\beta L|\right\},
\end{align}
where~$m$ and~$L$ are the curvature bounds from Assumption~\ref{ass:regularity}, and~$\alpha$ and~$\beta$ are the step sizes
of the prediction and correction steps in~\eqref{eq:prediction_iteration}
and~\eqref{eq:correction_iteration}, respectively. The terms~$\rho_p$
and~$\rho_c$ are the contraction factors of the prediction and correction
steps. Since~$\alpha, \beta \in \big(0, \frac{2}{L}\big)$, it follows from
Assumption~\ref{ass:regularity} that~$\rho_p, \rho_c \in
(0,1)$~\cite{SimonettoPCaTVcop2017}. For a general allocation, the following
proposition describes how the tracking error evolves from one sampling time
to the next.

\begin{proposition}[Prediction-Correction Tracking Error Bound~\cite{SimonettoPCaTVcop2017}] \label{prop:errorbound}
    Suppose Assumption~\ref{ass:regularity} holds, let~$\{(P_k,C_k)\}_{k=0}^{T-1}$ be any allocation, and let the iterates~$\{x_k\}_{k=0}^{T}$ be generated by the prediction and correction steps in~\eqref{eq:prediction_iteration}--\eqref{eq:correction_iteration}. Then, the following inequality holds:
    \begin{equation} \label{eq:error_recursion}
        e_{k+1}
        \leq
        a(P_k,C_k)e_k
        +
        d(P_k,C_k),\quad \forall k\in\{0,1,\dots,T-1\},
    \end{equation}
    where
    \begin{equation} \label{eq:contraction_factor}
    \begin{aligned}
        a(P_{k},C_{k})
        &:=
        \rho_c^{C_{k}}
        \left[
            \rho_p^{P_{k}}
            +
            \left(\rho_p^{P_{k}}+1\right)
            A_{\gamma}
        \right],\\
        d(P_{k},C_{k})
        &:=
        \frac{4h\Gamma}{m}
        \left(1+\frac{L}{m}\right)
        \rho_c^{C_{k}}
        \left(2\rho_p^{P_{k}}+1\right),\\
        A_{\gamma}
        &:=
        1-\gamma+\frac{2\gamma L}{m}.
    \end{aligned}
    \end{equation}    
\end{proposition}
\begin{proof}
    This result follows from Appendices A and C for case a) in~\cite{SimonettoPCaTVcop2017}. There, the bound is stated for a fixed number of prediction and correction steps. Because~\eqref{eq:error_recursion} relates only~$e_k$ and~$e_{k+1}$, the same argument applies at each~$k$ with~$(P_k,C_k)$ in place of the fixed counts.
\end{proof}

For ease of notation, let~$a_{j}:=a(P_{j},C_{j})$ and~$d_{j}:=d(P_{j},C_{j})$
for~$j\in\{0,1,\ldots,T-1\}$. Recursively
unrolling~\eqref{eq:error_recursion} gives
\begin{equation} \label{eq:unrolled_error_bound}
    e_k
    \leq
    \left(\prod_{j=0}^{k-1}a_{j}\right)e_0
    +
    \sum_{i=0}^{k-1}
    \left(\prod_{j=i+1}^{k-1}a_{j}\right)
    d_{i},\ \forall k\in\{0,1,\ldots,T\}.
\end{equation}
Moreover, since~$\rho_p, \rho_c \in (0,1)$, it follows
from~\eqref{eq:contraction_factor} that
\begin{equation}\label{eq:uniform_disturbance_bound}
    0\leq d_{j} \leq D_{0}h, \quad \forall j\in\{0,1,\ldots,T-1\},
\end{equation}
where~$D_{0}:=\frac{12\Gamma}{m}\big(1+\frac{L}{m}\big)$. The bound
in~\eqref{eq:unrolled_error_bound} holds for any allocation, and 
shows that both the initial
error~$e_0$ and each term~$d_i$ in the sum are multiplied by products of the
factors~$a_j$. If~$a_j < 1$ for all~$j\in\{0,1,\ldots,T-1\}$, then these
products decay geometrically in~$k$. The
factor~$a(P,C)$ in~\eqref{eq:contraction_factor} is nonincreasing in
both~$P$ and~$C$, and therefore ensuring~$a(P,C) < 1$ requires a
sufficient number of iterations per sample. Let us define the
minimum such number as
\begin{equation} \label{eq:Kmin}
    K_{\min}
    :=
    \min_{\substack{
        P,C \in \Z_{\geq0}\\
        a(P,C)<1
    }}
    P+C.
\end{equation}
The minimum in~\eqref{eq:Kmin} is attained, since~$a(0,C) < 1$ for
sufficiently large~$C$ and~$P+C\in\Z_{\geq 0}$. Since~$a(\cdot,\cdot)$ is
nonincreasing in each argument, any~$K \geq K_{\min}$ also admits a
pair~$(P,C)$ with~$P+C=K$ and~$a(P,C) < 1$.

For a given horizon~$T$ and budget~$B$, we define two regimes based
on~$K_{\min}$:
\begin{equation} \label{eq:Regimes}
    \n{Regime~I:}~\frac{B}{T} \geq K_{\min}~\n{ and }~\n{Regime~II:}~\frac{B}{T} < K_{\min}.
\end{equation}
Since the budget~$B$ is shared across all~$T$ samples, the regimes depend
on the average budget per sample~$B/T$. In Regime~I, the budget
allows~$P_k + C_k \geq K_{\min}$ for all~$k\in\{0,1,\ldots,T-1\}$. In
Regime~II, it does not, and therefore any allocation
satisfying~\eqref{eq:total_budget_constraint} has at least one~$k$
with~$a(P_{k},C_{k}) \geq 1$. The following subsections develop
allocations for each regime.

\subsection{Regime I: Sufficient Average Budget} \label{subsec:regime1_theory} 

Suppose that~$B/T\geq K_{\min}$, and define
\begin{equation} \label{eq:kbase&rB}
    K_{\n{base}}:= \left\lfloor \frac{B}{T}\right\rfloor,
\end{equation}
where~$K_{\n{base}}$ is the largest 
number of iterations
that makes all~$K_{k}$'s equal while obeying the budget
constraint in~\eqref{eq:total_budget_constraint}.
Since~$K_{\min}$ is an integer and~$B/T\geq K_{\min}$, we have~$K_{\text{base}}\geq K_{\min}$. Define $\mathcal{U}^{pc}:=\left\{(P,C)\in\Z_{\ge0}^{2}: P+C=K_{\n{base}},\, a(P,C)<1 \right\}$. We choose a pair
\begin{equation} \label{eq:regime1_allocation}
    (\bar P, \bar C) \in \mathcal{U}^{pc}, 
\end{equation}
and we set~$(P_k, C_k) = (\bar P, \bar C)$ for all~$k$ in~$\{0,1,\cdots,T-1\}$. Since~$K_{\n{base}} \geq K_{\min}$, the set~$\mathcal{U}^{pc}$ is nonempty. This uniform allocation uses~$TK_{\n{base}}\leq B$ total iterations
over the full time horizon, and therefore it satisfies~\eqref{eq:total_budget_constraint}.


\subsection{Regime II: Insufficient Average Budget} \label{subsec:regime2_theory} 
If~$B/T < K_{\min}$, then ~$a(P_{k},C_{k}) < 1$ cannot be guaranteed for
all~$k \in \{0,1,\ldots,T-1\}$. If the budget is sufficiently large, an allocation can still ensure that the products of the factors~$a(P_j,C_j)$ in~\eqref{eq:unrolled_error_bound} decay at a prescribed rate, which yields an explicit tracking bound over~$\{0, 1, \cdots, T\}$.
We formalize this notion with the following finite-horizon contraction condition.

\subsubsection{Finite-horizon contraction certificate} 
For user-specified choices of~$M\geq 1$ and~$\rho\in(0,1)$, we consider the condition
\begin{equation}\label{eq:finite_horizon_certificate}
    \prod_{j=i}^{k-1} a(P_{j},C_{j}) \leq M\rho^{k-i},
\end{equation}
for all~$i,k\in\Z_{\geq 0}$ with~$0\leq i<k\leq T$. We call~\eqref{eq:finite_horizon_certificate} a \emph{certificate} because,
when it holds, the tracking error can be shown to be bounded. This bound is
derived in Section~\ref{sec:pctracking}.

To enforce~\eqref{eq:finite_horizon_certificate} while choosing~$(P_k,C_k)$ one iteration at a time, we quantify the extent to which~$a(P_k,C_k)$ exceeds~$\rho$ at each iteration. 
Since~$\rho \in (0,1)$ and~$a(P,C) > 0$ for all choices of~$P \in \Z_{\ge 0}$ and~$C \in \Z_{\ge0}$, 
we define
\begin{equation} \label{eq:sdef}
    s(P_{k},C_{k}):= \log a(P_{k},C_{k}) - \log\rho.
\end{equation}
Intuitively,~$s(P_{k},C_{k})$ measures the extent to which a given pair~$(P_{k}, C_{k})$ 
ensures~$a(P_k,C_k) \leq \rho$ at iteration~$k$.
For any~$k$ in~$\{0,1,\ldots,T-1\}$, if~$s(P_{k}, C_{k}) \leq 0$, then~$a(P_k,C_k) \leq \rho$, and if~$s(P_{k}, C_{k}) > 0$,
then~$a(P_k,C_k) > \rho$. We also define
\begin{equation}
    q^{+}(q;P_{k},C_{k}) := \max\{0,\, q + s(P_{k},C_{k})\},
\end{equation}
which we use to measure the accumulated positive values of~$s(P_{k}, C_{k})$ and hence the cumulative failure of pairs~$(P_{k}, C_{k})$ to ensure~$a(P_k,C_k) \leq \rho$ over time.
Starting with~$q_{0}=0$, we set
\begin{equation}\label{eq:debt_recursion}
    q_{k+1}=q^{+}(q_{k};P_{k},C_{k}).
\end{equation}
The term~$q_{k}$ increases if~$a(P_{k},C_{k}) > \rho$ at iteration~$k$, decreases or remains at zero if~$a(P_{k},C_{k}) < \rho$ at iteration~$k$, and is reset to zero whenever the accumulated sum~$q + s(P_{k},C_{k})$ becomes negative at iteration~$k$. 
The following lemma shows that one can
enforce~\eqref{eq:finite_horizon_certificate}
by bounding~$q_{k}$ from~\eqref{eq:debt_recursion}.

\begin{lemma}[Necessary and sufficient condition for~\eqref{eq:finite_horizon_certificate}] \label{lem:debt_certificate}
  Let~$q_k$ be as defined in~\eqref{eq:debt_recursion}. Then the
  certificate~\eqref{eq:finite_horizon_certificate} holds for
  all~$i,k\in\Z_{\geq 0}$ satisfying~$0\leq i < k \leq T$ if and only if
  \begin{equation} \label{eq:debt_constraint}
      q_{k} \leq \log M \quad \n{for all }k \n{ in}~\{0,1,\ldots,T\}.
  \end{equation}
\end{lemma}
\begin{proof}
    \ifbool{short}{
    See authors' technical report~\cite{musavi2026computationallyconstrainedfirstorderpredictioncorrection}
    }{
    See Appendix~\ref{app:lem1}.
    }
\end{proof}

Lemma~\ref{lem:debt_certificate} provides a necessary and sufficient
condition for the certificate~\eqref{eq:finite_horizon_certificate} to hold. Since the
budget~$B$ is shared across iterations, maintaining this condition requires
reserving enough iterations for the remaining samples, which we quantify
next through the reserve function.

\subsubsection{Reserve function} \label{subsubsec:onlinealloc}
Let~$n\in\{0,1,\cdots,T-1\}$ denote the number of remaining samples 
of the objective function, 
and consider~$q\in[0,\log M]$. We define the \textit{reserve function} as
\begin{equation} \label{eq:reserve_dp}
    V_{n+1}(q):=
    \min_{\substack{(P,C)\in\Z_{\geq 0}^{2}\\ q^+(q;P,C)\leq\log M}}
    P+C+V_{n}\big(q^+(q;P,C)\big),
\end{equation}
where the constraint keeps~$q^+(q;P,C)\leq\log M$, and~$V_{n+1}(q)$ counts the minimum total number of prediction and correction iterations needed over the remaining~$n+1$ samples. 
We set $V_0(q):=0$, since no iterations are needed when no samples remain, and $V_{n+1}(q):=+\infty$ if no pair~$(P,C)$ satisfies the constraint in~\eqref{eq:reserve_dp}.
The value of~$V_{n}(q)$ also depends on the certificate parameters~$M$ 
and~$\rho$ from~\eqref{eq:finite_horizon_certificate} 
and the function~$a(P,C)$ from~\eqref{eq:contraction_factor},
and it is solved offline for all~$n$, before the horizon begins.
In practice, we compute~$V_n$ on a grid of possible values of~$q \in [0, \log M]$, rounding each transition~$q^+$ up to the next grid point, which can only increase~$V_n$ and therefore preserves the guarantees.

\subsubsection{Computing Allocations}
The reserve function determines how many iterations must be saved for
future samples so that~\eqref{eq:finite_horizon_certificate} can still be satisfied. It remains to choose~$P_{k}$ and~$C_{k}$ at each~$k$.
The remaining budget at step~$k$ for~$k\in\{0,1,\cdots,T\}$ is
\begin{equation} \label{eq:rkdef}
    R_{k}:=B-\sum_{j=0}^{k-1}(P_{j}+C_{j}).
\end{equation}
Then the feasible set for choosing~$P_k$ and~$C_k$
at iteration~$k$ is
\begin{multline} \label{eq:online_safe_set}
    U_k^{\mathrm{feasible}}
    :=
    \Big\{
    (P_{k},C_{k})\in\Z_{\geq 0}^{2}:
    q^+(q_k;P_{k},C_{k})\leq\log M,\\
    P_{k}+C_{k}+V_{T-k-1}\big(q^+(q_k;P_{k},C_{k})\big)\leq R_k
    \Big\},
\end{multline}
where the first condition ensures~$q_{k+1}\leq\log M$, so that, by
Lemma~\ref{lem:debt_certificate}, the
certificate~\eqref{eq:finite_horizon_certificate} continues to hold after
iteration~$k$, and the second 
condition ensures that the budget constraint in~\eqref{eq:total_budget_constraint} is satisfied. 

To choose among feasible pairs, we estimate the two terms
in~\eqref{eq:error_recursion} using information available at~$(x_k,t_k)$.
Since~$e_k$ cannot be evaluated online, we estimate it with
\begin{equation} \label{eq:online_error_bound}
    \widehat{e}_k:=\frac{\|\nabla_{x} f(x_k,t_k)\|}{m}.
\end{equation}
By strong convexity of~$f(\cdot,t_k)$, using the fact that the
solution~$x^{\star}(t_{k})$ satisfies~$\nabla_{x}f(x^{\star}(t_{k}),t_{k}) = 0$,
we find that~$\nabla_{x}f(x_{k},t_{k})^\top(x_{k}-x^{\star}(t_{k})) \geq me_{k}^{2}$.
Combined with the Cauchy--Schwarz inequality, this gives~$e_{k}\leq\widehat{e}_{k}$.
To estimate~$d(P,C)$ in~\eqref{eq:contraction_factor}, we define
\begin{equation} \label{eq:online_disturbance_score}
    \widehat d_k(P,C):=
    \frac{4h\|\nabla_{tx} f(x_k,t_k)\|}{m}
    \left(1+\frac{L}{m}\right)
    \rho_c^C(2\rho_p^P+1),
\end{equation}
which differs from~$d$ in~\eqref{eq:contraction_factor} by using the value
of~$\nabla_{tx}f$ at~$(x_k,t_k)$ instead of the worst-case bound~$\Gamma$
from Assumption~\ref{ass:regularity}.

At iteration~$k$, we then choose~$P_k$ and~$C_k$ using
\begin{equation} \label{eq:online_allocation_rule}
    (P_k,C_k)
    \in
    \argmin_{(P,C)\in U_k^{\mathrm{feasible}}} a(P,C)\widehat e_k+\widehat d_k(P,C).
\end{equation}
The objective in~\eqref{eq:online_allocation_rule} is a heuristic score that mimics the right-hand side of~\eqref{eq:error_recursion} using information available at~$t_k$. The procedure for allocating the budget in Regime~II is summarized in
Algorithm~\ref{alg:regime2}.

\begin{algorithm}[t]
\caption{Computationally Constrained Allocation}
\label{alg:regime2}
\begin{algorithmic}[1]
\REQUIRE Initial iterate~$x_0 \in \R^{n}$, horizon~$T \in \Z_{\geq 1}$,
budget $B \in \Z_{\geq 1}$ from~\eqref{eq:total_budget_constraint}, certificate parameters~$M \geq 1$ and~$\rho \in (0,1)$ from~\eqref{eq:finite_horizon_certificate},
and~$\{V_n\}_{n=0}^{T}$ from~\eqref{eq:reserve_dp}, computed offline
\STATE Verify $B\geq V_T(0)$
\STATE $q_0\gets0$, $R_0\gets B$
\FOR{$k=0,\cdots,T-1$}
    \STATE Compute~$\|\nabla_{tx}f(x_{k},t_{k})\|$ and~$\widehat{e}_k$ from~\eqref{eq:online_error_bound}
    \STATE Compute $\widehat d_k$ using
    \eqref{eq:online_disturbance_score}
    \STATE Form $U_k^{\mathrm{feasible}}$
    using \eqref{eq:online_safe_set}
    \STATE Select $(P_k,C_k)$ according to
    \eqref{eq:online_allocation_rule}
    \STATE Perform $P_k$ prediction iterations from~\eqref{eq:prediction_iteration}  using the objective~$f(\cdot,t_{k})$
    \STATE Sample the objective~$f(\cdot,t_{k+1})$
    \STATE Perform $C_k$ correction iterations from~\eqref{eq:correction_iteration} using the objective~$f(\cdot,t_{k+1})$ to obtain $x_{k+1}$
    \STATE $q_{k+1}\gets q^+(q_k;P_k,C_k)$
    \STATE $R_{k+1}\gets R_k-P_k-C_k$
\ENDFOR
\end{algorithmic}
\end{algorithm}

\section{Tracking Analysis} \label{sec:pctracking}

In this section, we analyze the allocation strategies derived for Regime~I in~\eqref{eq:regime1_allocation} and for Regime~II in~\eqref{eq:online_allocation_rule}.

\subsection{Regime I Tracking Guarantees}
With the allocation strategy from~\eqref{eq:regime1_allocation}, the tracking error of the prediction-correction algorithm that executes prediction steps using~\eqref{eq:prediction_iteration} and correction steps using~\eqref{eq:correction_iteration} follows directly from Proposition~\ref{prop:errorbound}.

\begin{corollary}[Tracking in Regime~I] \label{thm:regime1}
Suppose Assumption~\ref{ass:regularity} holds and~$B/T \geq K_{\min}$,
where~$B$ is from~\eqref{eq:total_budget_constraint} and~$K_{\min}$ is from~\eqref{eq:Kmin}.
Consider the allocation of prediction and correction steps 
$(\bar{P}, \bar{C})$ from~\eqref{eq:regime1_allocation}, and define~$\bar a:=a(\bar P,\bar C)$.
Then the budget constraint in~\eqref{eq:total_budget_constraint} is satisfied,~$\bar a<1$,
and we have
\begin{equation} \label{eq:regime1_tracking_bound}
    e_{k} \leq \bar{a}^{k} e_{0} + D_{0}h\,\frac{1-\bar{a}^{k}}{1-\bar{a}}
\end{equation}
for~$k\in\{0,1,\ldots,T\}$, where~$D_{0}$ is from~\eqref{eq:uniform_disturbance_bound}.
\end{corollary}
\begin{proof}
        The budget constraint holds by Section~\ref{subsec:regime1_theory},
    and~$\bar a<1$ because~$(\bar P,\bar C)\in\mathcal U^{pc}$. Since the
    allocation is constant,~$a_j=\bar a$ for all~$j$, and from~\eqref{eq:uniform_disturbance_bound},~$d_i\leq D_0h$ for all~$i$.
    Thus, the result follows from~\eqref{eq:unrolled_error_bound}. 
\end{proof}
Corollary~\ref{thm:regime1} shows that under Regime~I a uniform allocation suffices: the tracking error bound converges geometrically to a neighborhood of size~$D_0h/(1-\bar a)$. This neighborhood size 
depends on the sampling interval~$h$ from Section~\ref{subsec:tvop}, which can be changed by a user. 
We now bound the tracking error in Regime~II.

\subsection{Regime II Feasibility and Tracking Guarantees}
We first provide a sufficient condition under which~\eqref{eq:finite_horizon_certificate} can be satisfied.

\begin{theorem}[Certificate feasibility] \label{thm:exact-feasibility}
Suppose Assumption~\ref{ass:regularity} holds, and let~$M\geq 1$
and~$\rho\in(0,1)$ be given. If~$B \geq V_T(0)$, where~$V_T$ is
from~\eqref{eq:reserve_dp}, then there exists an allocation that satisfies
both the budget constraint~\eqref{eq:total_budget_constraint} and the
certificate~\eqref{eq:finite_horizon_certificate}. 
\end{theorem}
\begin{proof}
    \ifbool{short}{
    See authors' technical report~\cite{musavi2026computationallyconstrainedfirstorderpredictioncorrection}
    }{
    See Appendix~\ref{app:thm2}.
    }
\end{proof}

Under the condition of Theorem~\ref{thm:exact-feasibility},
Algorithm~\ref{alg:regime2} generates such an allocation, and its tracking
error is bounded as follows.

\begin{theorem}[Tracking in Regime~II] \label{thm:exact-tracking}
Suppose Assumption~\ref{ass:regularity} holds, we define
\begin{equation} \label{eq:thm3_ref1}
    \bar{d}:=\max_{0\leq j\leq T-1}d(P_j,C_j).
\end{equation}
and~$B \geq V_{T}(0)$ holds.
Then,~$U_{k}^{\mathrm{feasible}}$ is nonempty for
all~$k\in\{0,1,\ldots,T-1\}$. Moreover, the allocation generated by
Algorithm~\ref{alg:regime2} satisfies the budget constraint
in~\eqref{eq:total_budget_constraint}, and
\begin{equation}\label{eq:exact_tracking_bound}
    e_k \leq M\rho^{k}e_{0} + MD_0h\frac{1-\rho^k}{1-\rho},
    \quad \forall k\in \{0,1,\ldots,T\},
\end{equation}
where~$M$ and~$\rho$ are from~\eqref{eq:finite_horizon_certificate}
and~$D_{0}$ is from~\eqref{eq:uniform_disturbance_bound}.
\end{theorem}
\begin{proof}   
    \ifbool{short}{
    See authors' technical report~\cite{musavi2026computationallyconstrainedfirstorderpredictioncorrection}
    }{
    See Appendix~\ref{app:thm3}.
    }
\end{proof}

Theorem~\ref{thm:exact-tracking} provides a bound on the tracking error incurred by
allocations produced by Algorithm~\ref{alg:regime2}. 
As in Regime~I, the bound converges geometrically to a neighborhood whose
size is proportional to the sampling interval~$h$.

\subsubsection{A simpler condition using uniform allocation}
Checking the condition~$B \geq V_T(0)$ in Theorem~\ref{thm:exact-feasibility}
requires solving the dynamic program~\eqref{eq:reserve_dp}. A simpler
sufficient condition is obtained by restricting to uniform
allocations~$(P_k, C_k) = (P, C)$ for all~$k\in\{0,1,\ldots,T-1\}$, which
yields a budget condition, and an allocation that satisfies it, without
computing~$V_{n}$.

\begin{theorem}[Uniform-allocation budget condition]\label{thm:uniform-budget}
Suppose Assumption~\ref{ass:regularity} holds. Define
\begin{equation} \label{eq:uniform-budget}
    \theta_T:=\rho M^{1/T},
\end{equation}
where~$M\geq1$ and~$\rho\in(0,1)$ are from~\eqref{eq:finite_horizon_certificate}.
Also define 
\begin{equation}
    K_{\mathrm{unif}}
    :=
    \min_{\substack{P,C\in\mathbb Z_{\ge0}\\
    a(P,C)\leq\theta_T}}
    P+C,
\end{equation}
and let~$(P^\star,C^\star)$ be any pair such that~$P^\star + C^\star = K_{\mathrm{unif}}$. If $B\ge T\,K_{\mathrm{unif}}$, then the uniform allocation~$(P_k,C_k)\equiv(P^\star,C^\star)$ for~$k\in\{0,1,\cdots,T-1\}$ satisfies the budget constraint in~\eqref{eq:total_budget_constraint}, the certificate in~\eqref{eq:finite_horizon_certificate}, and the bound in~\eqref{eq:exact_tracking_bound}. Moreover,~$V_T(0)\leq B$, and therefore Algorithm~\ref{alg:regime2} can be used. 
\end{theorem}
\begin{proof}
    \ifbool{short}{
    See authors' technical report~\cite{musavi2026computationallyconstrainedfirstorderpredictioncorrection}
    }{
    See Appendix~\ref{app:thm4}.
    }
\end{proof}
Since~$V_T(0)\leq TK_{\mathrm{unif}}$, the condition in Theorem~\ref{thm:uniform-budget} is more conservative than that in Theorem~\ref{thm:exact-feasibility}, but it does not require computing~$V_n$.
The results in Theorem~\ref{thm:uniform-budget} show that we can still attain the bound in~\eqref{eq:exact_tracking_bound} even while using a uniform allocation of~$TK_{\n{unif}}$ total iterations. 

Moreover, since the allocation is the same for every~$k\in\{0,1,\ldots,T-1\}$,
computing~$K_{\mathrm{unif}}$ reduces to finding, for each number of
prediction steps~$P$, the minimum number of correction steps that
satisfies~$a(P,C)\leq\theta_T$. Since~$\rho_c\in(0,1)$, solving this
inequality for~$C$ gives
\begin{equation}
    C_{\min}(P) :=
    \max\left\{
        0,\
        \left\lceil
        \frac{ \log\!\left( \theta_T/ [A_\gamma+(1+A_\gamma)\rho_p^P] \right)}{\log\rho_c}
        \right\rceil
    \right\},
\end{equation}
where~$A_{\gamma}$ is from~\eqref{eq:contraction_factor}, which implies that
\begin{equation}
    K_{\mathrm{unif}}
    =
    \min_{\substack{P\in\mathbb Z_{\ge0} \\ a(P,C_{\min}(P)) \leq \theta_{T}}}
    P+C_{\min}(P).
\end{equation}
Then the uniform allocation calculation is reduced to an integer search over prediction counts, 
without using a dynamic program. 
\textcolor{gray}{
}
\section{Numerical Experiments} \label{sec:simulations}
In this section, we compare our adaptive allocation against a uniform allocation under the same total iteration budget~$B$.

\subsection{Problem and Experimental Setup}
\label{subsec:simulation_setup}
Consider the following unconstrained problem, adapted
from~\cite{SimonettoPCaTVcop2017}, in which the minimizer follows a moving
reference whose speed changes over the horizon:
\begin{equation}
\label{eq:simulation_objective}
\begin{aligned}
    \min_{x\in\mathbb R^2}\quad
    f(x,t)
    :={}&
    \frac12(x-r(t))^\top Q(x-r(t))\\
    &+\kappa\log\!\left(1+\exp(\mu c^\top x)\right),
\end{aligned}
\end{equation}
where $Q\succ0$, $c \in \R^{2}$ is a unit vector, $\kappa,\mu>0$, and the reference~$r(t)=(\cos\phi(t),\sin\phi(t))^\top$ traces a closed curve with angular speed
\begin{equation} \label{eq:simulation_phase}
    \dot\phi(t)
    :=\omega_{\mathrm{s}}
    +\frac{\omega_{\mathrm{f}}-\omega_{\mathrm{s}}}{2}
    \left[
        \tanh\!\left(\frac{t-t_{\mathrm{s}}}{w}\right)
        -
        \tanh\!\left(\frac{t-t_{\mathrm{e}}}{w}\right)
    \right],
\end{equation}
which produces a slow--fast--slow pattern of temporal variation.
The objective in~\eqref{eq:simulation_objective} satisfies Assumption~\ref{ass:regularity} since $\nabla_{xx}f$ is uniformly bounded between $mI$ and $LI$ with~$m=\lambda_{\min}(Q)$ and~$L=\lambda_{\max}(Q)+\kappa\mu^2/4$, and since $\nabla_{tx}f(x,t)=-Q\dot r(t)$ gives $\|\nabla_{tx}f(x,t)\|\leq \|Q\|_2\sup_t|\dot\phi(t)|=:\Gamma$.
The optimizer $x^\star(t_k)$ is computed numerically and is used only to evaluate tracking performance.

We use~$Q= \big[ \begin{smallmatrix} 1 & 0 \\ 0 & 2 \end{smallmatrix} \big]$,~$c=(1,1)^\top/\sqrt{2}$,~$\kappa=\mu=1$,~$\omega_{\mathrm{s}}=0.05$,~$\omega_{\mathrm{f}}=1$,~$t_{\mathrm{s}}=3.6$,~$t_{\mathrm{e}}=9.6$, and~$w=0.5$, which gives $m=1$, $L=2.25$, and $\Gamma=2$.
We also use $T=20$, $h=1.2$, $\alpha = 2/(m+L)$, $\beta=0.2$, $\gamma=0.5$, $B=140$, $M=1000$, and $\rho = 0.95$. The step size~$\alpha=2/(m+L)$ minimizes~$\rho_p$, and the remaining
parameters are chosen so that the fast phase of~\eqref{eq:simulation_phase}
occurs within the horizon and the experiment lies in Regime~II.
Here~$K_{\min}=8$ while~$B/T=7$, and therefore the experiment lies in Regime II. For the adaptive allocation, we compute the reserve function in~\eqref{eq:reserve_dp} on a uniform grid of~$q$ values with spacing~$0.03$.
The uniform baseline uses~$(P_k,C_k)=(2,5)$ for all~$k$. Both methods use~$x_0=(3,-3)^\top$, which gives~$e_0\approx 3.69$, and we set~$\phi(0)=0$.
To compare the two methods, we use the cumulative tracking errors
\begin{equation}
\label{eq:simulation_metrics}
    E_T:=\sum_{k=1}^{T}e_k
    \quad\text{and}\quad
    e_{\max}:=\max_{1\leq k\leq T}e_k,
\end{equation}
where $e_k$ is from~\eqref{eq:tracking_error}, and we refer to~$e_T$ as the terminal error. The initialization error~$e_0$ is excluded from~$E_T$ and~$e_{\max}$ because the prediction-correction method cannot change it.

\subsection{Adaptive and Uniform Allocation}
\label{subsec:simulation_comparison}
Under the above conditions, adaptive allocation reduces the cumulative tracking error~$E_{T}$ by about~$45.5\%$ and the maximum error~$e_{\max}$ by about~$64.7\%$ relative to the uniform baseline, at the cost of a larger terminal error ($0.0132$ for the adaptive allocation vs.~$0.00177$ for the uniform baseline).
This behavior follows from the online allocation rule in~\eqref{eq:online_allocation_rule}. Since~$a(P, C)$ and~$\widehat d_k(P, C)$ are nonincreasing in~$P$
and~$C$, the rule spends as many iterations as~$U_k^{\mathrm{feasible}}$
allows. As a result, the adaptive allocation front-loads the budget: it
uses~$10$ iterations per interval during the first seven intervals, while
reserving enough budget to satisfy the certificate, and it then tapers to
two prediction steps per interval over the final five intervals. Spending
more iterations early reduces the large initial error quickly, which
lowers~$E_T$ and~$e_{\max}$, while the small allocation at the end leads to
the larger terminal error. The uniform allocation uses~$(2,5)$ throughout. 

\begin{figure}[t]
    \centering
    \includegraphics[width=0.8\linewidth]{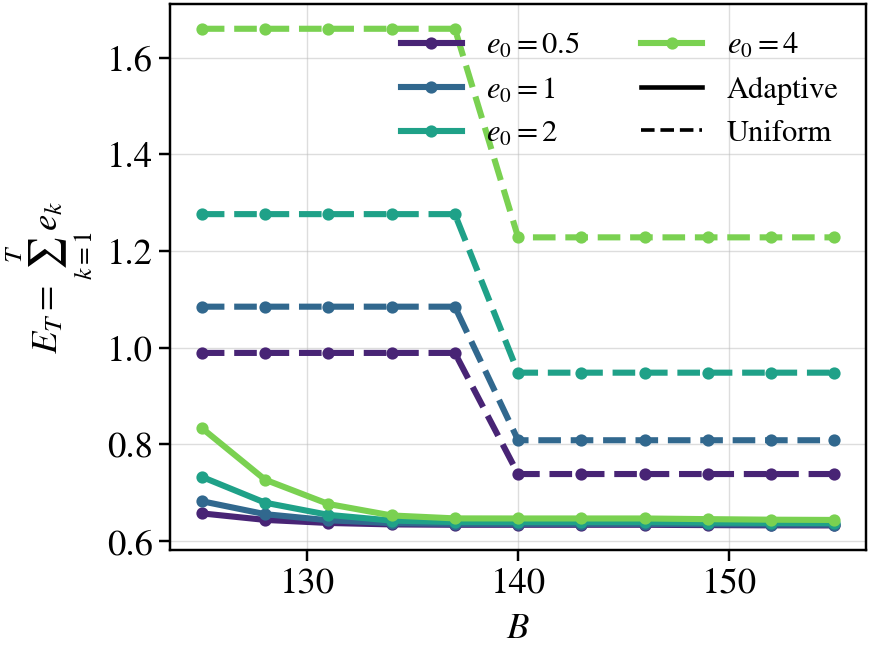}
    \vspace{-1.5em}
    \caption{
    Cumulative error vs. budget $B$, adaptive (solid) against uniform (dashed), at four values of~$e_0$.
    The drop in the uniform method between~$B=137$ and~$B=140$ is due to the increase in~$\lfloor B/T\rfloor$ from~$6$ to~$7$. The adaptive
    allocation outperforms the uniform allocation
    for all budgets and initial errors considered. 
    }
    \label{fig:budget-sensitivity}
\end{figure}

\begin{figure}[t]
    \centering
    \includegraphics[width=0.8\linewidth]{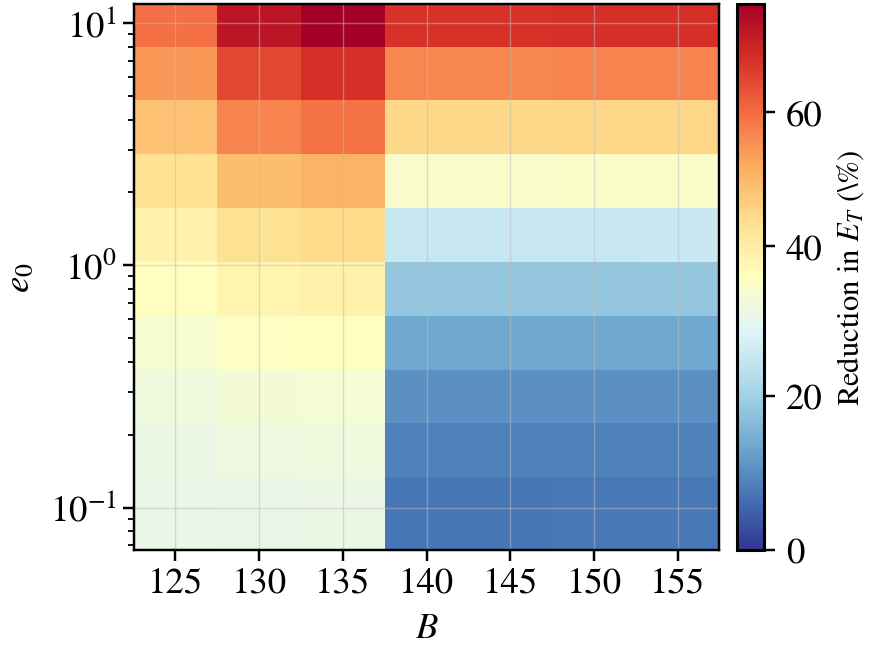}
    \vspace{-1.5em}
    \caption{
    Reduction in cumulative tracking error over the joint~$(B,e_0)$ grid.
    The adaptive method provides the largest decreases in error for large~$e_{0}$ and for budgets below~$B=140$, and it decreases the error for all values of these parameters considered. 
    }
    \label{fig:budget-init-grid}
    \vspace{-1em}
\end{figure}

\subsection{Sensitivity to Initialization and Budget}
\label{subsec:sensitivity}
This subsection examines whether the advantage of the adaptive allocation
persists across initial errors and budgets.
We first repeat the preceding experiment with $B=140$ for ten initial-error magnitudes~$\eta_j=10^{-1+2j/9}$, $j=0,\cdots,9$, holding the direction of~$x_0-x^\star(0)$ fixed.
The reduction in~$E_{T}$ grows with~$\eta_j$, from about~$7.4\%$ at~$\eta_0=0.1$ to about~$67.7\%$ at~$\eta_9=10$.
The terminal error of each method is the same across this range ($0.013$ for the adaptive allocation and~$0.002$ for the uniform baseline), and therefore terminal error alone does not reflect how well either method tracked the trajectory.

Next, we examine the cumulative error between the uniform baseline and the adaptive allocation with the budget~$B$ ranging over~$\{125,128,\ldots,155\}$.
These values all lie above the value~$V_T(0)=122$ and below the Regime I budget~$TK_{\min}=160$. We consider the initial error values~$e_0\in\{0.5,1,2,4\}$, which are shown in Figure~\ref{fig:budget-sensitivity}. 
The adaptive allocation achieves a lower cumulative tracking error for all
budgets and initial errors considered.

Finally, we examine the error reduction between the adaptive allocation 
and the uniform baseline
with a budget~$B$ ranging over~$\{125,130,\ldots,155\}$ and values of the initial error ranging over~$e_{0} \in \{\eta_{0},\eta_{1},\ldots,\eta_{9}\}$. 
We find that the error reduction ranges from~$7.4\%$ ($B=140$, $e_0=\eta_{0}$) to~$76.1\%$ ($B=135$, $e_0=\eta_{9}$), with an average of~$35.4\%$, which is shown in Figure~\ref{fig:budget-init-grid}.
The decrease in this reduction 
that occurs 
between a budget of~$135$ and~$140$ corresponds to the improvement in the cumulative tracking error of the uniform baseline due to the increase in~$\lfloor B/T\rfloor$ from~$6$ (with an allocation of~$(2,4)$) to~$7$ (with an allocation of~$(2,5)$) iterations.
The advantage of the adaptive method is largest for large initial errors and for budgets below~$B=140$ where the uniform baseline is limited to six iterations per sample, and the adaptive method
produces a decrease in error across all values of the budget and initial error.

\section{Conclusion} \label{sec:conclusion}

This paper presented a method of adaptive allocation for the number of prediction and correction steps 
in a prediction-correction algorithm that is subject to computational constraints. 
We showed that 
when the average computational budget per sample exceeds a certain threshold, a uniform allocation suffices for a decaying tracking error bound. 
When it does not, we introduced a finite-horizon contraction certificate that yields budget conditions and an allocation algorithm that enforce explicit tracking error bounds. 
Simulations show that the adaptive allocation reduces cumulative tracking error compared with a uniform allocation across all problems considered. 
Future work includes extending this approach to other prediction schemes.


\red{
}

\bibliographystyle{IEEEtran}
\bibliography{references}

\appendix
\subsection{Proof of Lemma~\ref{lem:debt_certificate}} \label{app:lem1}
        The recursion in~\eqref{eq:debt_recursion} gives
        \begin{equation}
        q_k=
        \max\left\{
            0,\
            \max_{0\leq i<k}
            \sum_{j=i}^{k-1}s(P_j,C_j)
        \right\}.    
        \end{equation}
        Consequently, using~\eqref{eq:sdef}, we see that
        having $q_k\leq\log M$ for every $k$ in~$\{0,1,\cdots,T\}$
        is equivalent to having
            \begin{equation}
            \sum_{j=i}^{k-1}
            \bigl(\log a(P_j,C_j)-\log\rho\bigr)
            \leq\log M
            \end{equation}
        for every~$i\in\Z_{\geq 0}$ satisfying~$0\leq i<k\leq T$.
        Exponentiating proves the equivalence.

\subsection{Proof of Theorem~\ref{thm:exact-feasibility}} \label{app:thm2}
    Suppose $V_T(0)\leq B$. We
    construct a feasible allocation using the minimizers in~
    \eqref{eq:reserve_dp}. 
    Since $V_T(0)\leq B<+\infty$ and the values
    of~$V_{n}$ from~\eqref{eq:reserve_dp} are nonnegative integers or $+\infty$, the minimum defining~$V_T(0)$ is attained. Let $(P_0,C_0)$ be a minimizer and ~$q_1=q^+(0;P_0,C_0)$, so that 
    \begin{equation}
        V_T(0) = (P_0+C_0)+V_{T-1}(q_1),
    \end{equation}
    which implies~$V_{T-1}(q_1)<+\infty$. 
    Repeating the same argument at
    $q_1,q_2,\ldots$ where
    $q_{k+1}=q^+(q_k;P_k,C_k)$ is from~\eqref{eq:debt_recursion} for $k=0,\ldots,T-1$, yields
    \begin{equation}
        \begin{aligned}
            V_T(0)
            &= \sum_{k=0}^{T-1}(P_k+C_k)+V_0(q_T)
            = \sum_{k=0}^{T-1}(P_k+C_k).
        \end{aligned}
    \end{equation}
    This allocation satisfies the budget constraint in~\eqref{eq:total_budget_constraint} since~$\sum_{k=0}^{T-1}(P_k+C_k)=V_T(0)\leq B$. Moreover, each minimizer satisfies the constraint $q_{k+1} \leq \log M$ in~\eqref{eq:reserve_dp}, so the certificate~\eqref{eq:finite_horizon_certificate} holds by Lemma~\ref{lem:debt_certificate}.

\subsection{Proof of Theorem~\ref{thm:exact-tracking}} \label{app:thm3}

    We show by induction that $R_{k} \geq V_{T-k}(q_{k})$ and~$U_{k}^{\n{feasible}}\neq\emptyset$ for all~$k\in\{0,1,\ldots,T-1\}$. For~$k=0$, we have~$R_0=B\geq V_T(0)=V_T(q_0)$. Suppose~$R_k\geq V_{T-k}(q_k)$. Then~$V_{T-k}(q_k)<+\infty$, and the minimum in~\eqref{eq:reserve_dp} defining~$V_{T-k}(q_k)$ is attained by some pair~$(\bar P,\bar C)$. This pair satisfies~$q^+(q_k;\bar P,\bar C)\leq\log M$ and
    \begin{equation*}
        \bar P+\bar C+V_{T-k-1}\big(q^+(q_k;\bar P,\bar C)\big)=V_{T-k}(q_k)\leq R_k.
    \end{equation*}
    Then~$(\bar P,\bar C)\in U_k^{\mathrm{feasible}}$. For the selected~$(P_k,C_k)\in U_k^{\mathrm{feasible}}$, the second condition in~\eqref{eq:online_safe_set} gives~$R_{k+1}=R_k-P_k-C_k\geq V_{T-k-1}(q_{k+1})$, which completes the induction.

    Since~$V_0\equiv 0$, we have~$R_T\geq 0$, and~\eqref{eq:total_budget_constraint} holds. The first condition in~\eqref{eq:online_safe_set} gives~$q_k\leq\log M$ for all~$k\in\{0,1,\ldots,T\}$, and the certificate in~\eqref{eq:finite_horizon_certificate} holds by Lemma~\ref{lem:debt_certificate}.
    Using the tracking error recursion in~\eqref{eq:unrolled_error_bound} and the certificate in~\eqref{eq:finite_horizon_certificate}, where an empty product equals~$1\leq M$, we find 
    \begin{equation} \label{eq:thm3_step1}
        e_k \leq M\rho^{\,k}e_0 + M\sum_{i=0}^{k-1} \rho^{k-i-1} d_{i}.
    \end{equation}
    Since~$d_{i}\leq \bar{d}$ from~\eqref{eq:thm3_ref1}, we find
    \begin{equation} \label{eq:thm3_step2}
    e_k \leq M\rho^{\,k}e_0 + M \bar{d}\sum_{i=0}^{k-1}\rho^{k-i-1}
    = M\rho^{\,k}e_0 + M\bar{d}\frac{1-\rho^{k}}{1-\rho}.
    \end{equation}
    The result follows from bounding~$\bar{d}\leq D_0h$ using~\eqref{eq:uniform_disturbance_bound}.

    

\subsection{Proof of Theorem~\ref{thm:uniform-budget}} \label{app:thm4}
    Consider the uniform allocation~$(P_k,C_k)\equiv(P^\star,C^\star)$ for~$k\in\{0,1,\ldots,T-1\}$. Since~$P^\star+C^\star=K_{\mathrm{unif}}$, the total budget satisfies~$\sum_{k=0}^{T-1}(P_k+C_k) = T K_{\mathrm{unif}}\leq B$.
    Moreover, by the definition of~$K_{\mathrm{unif}}$, we have~$a(P^\star,C^\star)\leq\theta_T=\rho M^{1/T}$. Therefore, for all integers~$i$ and~$k$ such that~$0 \leq i < k \leq T$, we have
    \begin{equation*}
    \prod_{j=i}^{k-1}a(P_j,C_j)
    =
    a(P^\star,C^\star)^{k-i}
    \leq
    \rho^{k-i}M^{(k-i)/T} \leq
    M\rho^{k-i},
    \end{equation*}
    where the last inequality follows from~$M\ge1$ and~$(k-i)/T\leq1$. Hence, the certificate in~\eqref{eq:finite_horizon_certificate} is satisfied.
    Following the same steps from~\eqref{eq:thm3_step1} to~\eqref{eq:thm3_step2} in the proof of Theorem~\ref{thm:exact-tracking}, we find
    \begin{equation*}
        e_k \leq M\rho^{\,k}e_0 + MD_{0}h\frac{1-\rho^{k}}{1-\rho}.
    \end{equation*}
    Finally, by Lemma~\ref{lem:debt_certificate}, the uniform allocation keeps~$q_k\leq\log M$ for all~$k$, and therefore it is feasible in~\eqref{eq:reserve_dp} at every stage. Hence~$V_T(0)\leq TK_{\mathrm{unif}}\leq B$.


\end{document}